\documentstyle[12pt]{article}
\begin{document}

\def\ds{\displaystyle}
\def\beq{\begin{equation}}
\def\eeq{\end{equation}}
\def\bea{\begin{eqnarray}}
\def\eea{\end{eqnarray}}
\def\beeq{\begin{eqnarray}}
\def\eeeq{\end{eqnarray}}
\def\ve{\vert}
\def\vel{\left|}
\def\ver{\right|}
\def\nnb{\nonumber}
\def\ga{\left(}
\def\dr{\right)}
\def\aga{\left\{}
\def\adr{\right\}}
\def\lla{\left<}
\def\rra{\right>}
\def\rar{\rightarrow}
\def\nnb{\nonumber}
\def\la{\langle}
\def\ra{\rangle}
\def\ba{\begin{array}}
\def\ea{\end{array}}
\def\tr{\mbox{Tr}}
\def\ssp{{\Sigma^{*+}}}
\def\sso{{\Sigma^{*0}}}
\def\ssm{{\Sigma^{*-}}}
\def\xis0{{\Xi^{*0}}}
\def\xism{{\Xi^{*-}}}
\def\qs{\la \bar s s \ra}
\def\qu{\la \bar u u \ra}
\def\qd{\la \bar d d \ra}
\def\qq{\la \bar q q \ra}
\def\gGgG{\la g^2 G^2 \ra}
\def\q{\gamma_5 \not\!q}
\def\x{\gamma_5 \not\!x}
\def\g5{\gamma_5}
\def\sb{S_Q^{cf}}
\def\sd{S_d^{be}}
\def\su{S_u^{ad}}
\def\ss{S_s^{??}}
\def\sbp{{S}_Q^{'cf}}
\def\sdp{{S}_d^{'be}}
\def\sup{{S}_u^{'ad}}
\def\ssp{{S}_s^{'??}}
\def\sig{\sigma_{\mu \nu} \gamma_5 p^\mu q^\nu}
\def\fo{f_0(\frac{s_0}{M^2})}
\def\ffi{f_1(\frac{s_0}{M^2})}
\def\fii{f_2(\frac{s_0}{M^2})}
\def\O{{\cal O}}
\def\sl{{\Sigma^0 \Lambda}}
\def\es{\!\!\! &=& \!\!\!}
\def\ar{&+& \!\!\!}
\def\ek{&-& \!\!\!}
\def\cp{&\times& \!\!\!}
\def\se{\!\!\! &\simeq& \!\!\!}
\def\kpm{&\pm& \!\!\!}
\def\kmp{&\mp& \!\!\!}

% .........................................................

\def\simlt{\stackrel{<}{{}_\sim}}
\def\simgt{\stackrel{>}{{}_\sim}}

% .........................................................

\title{
         {\Large
                 {\bf
T violation in $\Lambda_b \rar \Lambda \ell^+ \ell^-$ decay 
beyond standard model
                 }
         }
      }

\author{\vspace{1cm}\\
{\small T. M. Aliev$^a$ \thanks
{e-mail: taliev@metu.edu.tr}\,\,,
A. \"{O}zpineci$^b$ \thanks
{e-mail: ozpineci@ictp.trieste.it}\,\,,
M. Savc{\i}$^a$ \thanks
{e-mail: savci@metu.edu.tr} \,\,,
C. Y\"{u}ce$^a$}\\
{\small a Physics Department, Middle East Technical University, 
06531 Ankara, Turkey}\\
{\small b  The Abdus Salam International Center for Theoretical Physics,
I-34100, Trieste, Italy} }
\date{}

\begin{titlepage}
\maketitle
\thispagestyle{empty}

\begin{abstract}
Using the most general, model independent form of effective Hamiltonian,
the transverse polarization $P_T$ of $\Lambda$ in $\Lambda_b \rar \Lambda
\ell^+ \ell^-$ decay is studied. It is observed that the averaged 
$\lla P_T \rra$ is very sensitive to the existence of new physics and 
can attain large values, which can be measured at future colliders.
\end{abstract}

%\vspace{1cm}
~~~PACS numbers: 12.60.--i, 13.30.--a, 14.20.Mr
\end{titlepage}

\section{Introduction}
Recently time--reversal (T) violation has been measured in the $K^0$ system
\cite{R4801}. However, the origin of T violation, as well as CP violation
which also has been obtained experimentally in $K^0$ meson system,
remains unclear. In the standard model (SM) both violations comes from a
phase of the Cabibbo-Kobayashi--Maskawa (CKM) matrix \cite{R4802}. It is
well known that SM predicts sizeable CP violation in the $B^0$ meson system
(see \cite{R4803}). Detecting CP violation constitutes one of the main research
directions of the working B meson factories \cite{R4804}. These machines
have already provided the first evidence for CP violation in B system, namely 
$\sin 2\beta=0.78 \pm 0.08$ \cite{R4805}. In search of other sources of CP
violation one needs to investigate new processes. In this work we study the
T--violating effects in the baryonic $\Lambda_b \rar \Lambda \ell^+ \ell^-$
($\ell = \mu,~\tau$) 
decays using the most general form of the effective Hamiltonian. Large
number of $\Lambda_b$ baryons are expected to be produced in Tevatron, LHC,
etc., which is a great opportunity to examine the SM predictions for various
decay modes of $\Lambda_b$. These decays are also very sensitive to the new
physics beyond the SM. The interest to baryonic decays can be attributed to 
the fact that only these decays can give valuable information about the 
handedness of the quarks.  
It should be noted that T--violating effects in the $\Lambda_b \rar \Lambda
\ell^+ \ell^-$ and $B \rar K^\ast \ell^+ \ell^-$ decays were studied in the 
Supersymmetric model in \cite{R4806} and \cite{R4807}, respectively.  

It is well known that for a general three--body decay the triple
spin--momentum correlations $\vec{s}\cdot (\vec{p}_i\times\vec{p}_j)$ are
T--odd observables, where $\vec{s},~\vec{p}_i$ and $\vec{p}_j$ are the spin
and momenta of final particles. So, in order to organize T--odd quantities we
have two possibilities:
\begin{itemize}
\item either we choose lepton polarization as the polarization of final 
particles, or
\item we choose only $\Lambda$ baryon polarization.
\end{itemize}
 
In the $\Lambda_b \rar \Lambda \ell^+ \ell^-$ decay, experimentally only one
polarization state of the particles can be measured. In principle one can
use either polarization of the lepton or polarization of the baryon in
studying T--odd observables. The first possibility , i.e., the lepton
polarizations in the $\Lambda_b \rar \Lambda \ell^+ \ell^-$ decay was
studied in detail in \cite{R4808}. Therefore, in the present work we prefer 
choose the second possibility, namely, $\Lambda$ baryon polarization in
investigating the T--violating effects. It should be noted here that, lepton
polarization effects for the $B \rar K^\ast \ell^+ \ell^-$ were studied
comprehensively in \cite{R4809}.  

The paper is organized as follows. In section
2, using the most general, model independent form of the effective
Hamiltonian, we derive the expression for the T--odd transverse polarization
of $\Lambda$. Section 3 is devoted to the numerical analysis and 
concluding remarks.

\section{Calculation of the transversal polarization of $\Lambda$ baryon}

The matrix element of the $\Lambda_b \rar\Lambda \ell^+\ell^-$ decay at
quark level is described by the $b \rar s \ell^+\ell^-$ transition. The
decay amplitude for the $b \rar s \ell^+\ell^-$ transition, in a general
model independent form, can be written in the following way (see
\cite{R4810,R4811})

\bea
\label{e1}
\lefteqn{
{\cal M} = \frac{G \alpha}{\sqrt{2} \pi} V_{tb}V_{ts}^\ast \Bigg\{
C_{SL} \bar s i \sigma_{\mu\nu} \frac{q^\nu}{q^2} L b \bar\ell \gamma^\mu \ell +
C_{BR} \bar s i \sigma_{\mu\nu} \frac{q^\nu}{q^2} b \bar\ell \gamma^\mu \ell +
C_{LL}^{tot} \bar s_L \gamma^\mu b_L \bar \ell_L\gamma_\mu \ell_L} \nnb\\
\ar C_{LR}^{tot} \bar s_L \gamma^\mu b_L \bar \ell_R \gamma_\mu \ell_R +
C_{RL} \bar s_R \gamma^\mu b_R \bar \ell_L \gamma_\mu \ell_L +
C_{RR} \bar s_R \gamma^\mu b_R \bar \ell_R \gamma_\mu \ell_R \nnb \\
\ar C_{LRLR} \bar s_L b_R \bar \ell_L \ell_R +
C_{RLLR} \bar s_R b_L \bar \ell_L \ell_R +
C_{LRRL} \bar s_L b_R \bar \ell_R \ell_L +
C_{RLRL} \bar s_R b_L \bar \ell_R \ell_L \nnb \\
\ar C_T \bar s \sigma^{\mu\nu} b \bar \ell \sigma_{\mu\nu} \ell +
i C_{TE} \epsilon^{\mu\nu\alpha\beta} \bar s \sigma_{\mu\nu} s
\sigma_{\alpha\beta} \ell \Bigg\}~,
\eea
where $L=(1-\gamma_5)/2$ and $R=(1+\gamma_5)/2$ are the chiral operators and
$C_X$ are the coefficients of the four--Fermi interaction. Part of these
Wilson coefficients and structures does already exist in the effective
Hamiltonian of the $b \rar s$ transition in the SM. 
The first two of the coefficients $C_{SL}$ and 
$C_{BR}$ are the nonlocal Fermi interactions which correspond to $-2 m_s
C_7^{eff}$ and $-2 m_b C_7^{eff}$ in the SM, respectively. The following
four terms describe vector type interactions. Two of these vector
interactions containing the coefficients $C_{LL}^{tot}$ and $C_{LR}^{tot}$ do
also exist in the SM in the forms $(C_9^{eff}-C_{10})$ and
$(C_9^{eff}+C_{10})$, respectively. Therefore $C_{LL}^{tot}$ and
$C_{LR}^{tot}$ represent the sum of the combinations from SM and the new
physics in the following forms 
\bea
\label{e2}
C_{LL}^{tot} \es C_9^{eff}- C_{10} + C_{LL}~, \nnb \\
C_{LR}^{tot} \es C_9^{eff}+ C_{10} + C_{LR}~.
\eea
The terms with  $C_{LRRL},~C_{LRLR},~C_{RLRL}$ and $C_{RLLR}$ describe the
scalar type interactions. The last two terms in Eq. (\ref{e1}) correspond to
the tensor type interactions. The amplitude of the exclusive $\Lambda_b
\rar\Lambda \ell^+\ell^-$ decay can be obtained by sandwiching the matrix
element of the $b \rar s \ell^+ \ell^-$ decay between initial and final
state baryons. It follows from Eq. (\ref{e1}) that, in order to calculate
the amplitude of the $\Lambda_b \rar\Lambda \ell^+\ell^-$ decay the
following matrix elements are needed 
\bea
\label{e3}
&&\lla \Lambda \vel \bar s \gamma_\mu (1 \mp \gamma_5) b \ver \Lambda_b
\rra~,\nnb \\
&&\lla \Lambda \vel \bar s \sigma_{\mu\nu} (1 \mp \gamma_5) b \ver \Lambda_b
\rra~,\nnb \\
&&\lla \Lambda \vel \bar s (1 \mp \gamma_5) b \ver \Lambda_b \rra~.
\eea
Explicit forms of these matrix elements in terms of the form factors are
presented in Appendix--A of \cite{R4812}. Using the parametrization of these 
matrix elements, the matrix form of the $\Lambda_b \rar\Lambda \ell^+\ell^-$ 
decay can be written as 
\bea
\label{e4}
\lefteqn{
{\cal M} = \frac{G \alpha}{4 \sqrt{2}\pi} V_{tb}V_{ts}^\ast \Bigg\{
\bar \ell \gamma^\mu \ell \, \bar u_\Lambda \Big[ A_1 \gamma_\mu (1+\gamma_5) +
B_1 \gamma_\mu (1-\gamma_5) }\nnb \\
\ar i \sigma_{\mu\nu} q^\nu \big[ A_2 (1+\gamma_5) + B_2 (1-\gamma_5) \big]
+q_\mu \big[ A_3 (1+\gamma_5) + B_3 (1-\gamma_5) \big]\Big] u_{\Lambda_b}
\nnb \\
\ar \bar \ell \gamma^\mu \gamma_5 \ell \, \bar u_\Lambda \Big[
D_1 \gamma_\mu (1+\gamma_5) + E_1 \gamma_\mu (1-\gamma_5) +
i \sigma_{\mu\nu} q^\nu \big[ D_2 (1+\gamma_5) + E_2 (1-\gamma_5) \big]
\nnb \\
\ar q_\mu \big[ D_3 (1+\gamma_5) + E_3 (1-\gamma_5) \big]\Big] u_{\Lambda_b}+
\bar \ell \ell\, \bar u_\Lambda \big(N_1 + H_1 \gamma_5\big) u_{\Lambda_b}
+\bar \ell \gamma_5 \ell \, \bar u_\Lambda \big(N_2 + H_2 \gamma_5\big) 
u_{\Lambda_b}\nnb \\
\ar 4 C_T \bar \ell \sigma^{\mu\nu}\ell \, \bar u_\Lambda \Big[ f_T 
\sigma_{\mu\nu} - i f_T^V \big( q_\nu \gamma_\mu - q_\mu \gamma_\nu \big) -
i f_T^S \big( P_\mu q_\nu - P_\nu q_\mu \big) \Big] u_{\Lambda_b}\nnb \\
\ar 4 C_{TE} \epsilon^{\mu\nu\alpha\beta} \bar \ell \sigma_{\alpha\beta}
\ell \, i \bar u_\Lambda \Big[ f_T \sigma_{\mu\nu} - 
i f_T^V \big( q_\nu \gamma_\mu - q_\mu \gamma_\nu \big) -
i f_T^S \big( P_\mu q_\nu - P_\nu q_\mu \big) \Big] u_{\Lambda_b}\Bigg\}~,
\eea
where $P=p_{\Lambda_b}+ p_\Lambda$.

Explicit expressions of the functions $A_i,~B_i,~D_i,~E_i,~H_j$ and $N_j$
$(i=1,2,3$ and $j=1,2)$ are given in Appendix--A of \cite{R4812}. 

Obviously, the $\Lambda_b \rar\Lambda \ell^+\ell^-$ decay introduces a lot
of form factors. However, when the heavy quark effective theory (HQET) has
been used, the heavy quark symmetry reduces the number of independent form
factors to two only $(F_1$ and $F_2)$, irrelevant with the Dirac structure of
the relevant operators \cite{R4813}, and hence we obtain that
\bea
\label{e5}
\lla \Lambda(p_\Lambda) \vel \bar s \Gamma b \ver \Lambda(p_{\Lambda_b})
\rra = \bar u_\Lambda \Big[F_1(q^2) + \not\!v F_2(q^2)\Big] \Gamma
u_{\Lambda_b}~,
\eea
where $\Gamma$ is an arbitrary Dirac structure,
$v^\mu=p_{\Lambda_b}^\mu/m_{\Lambda_b}$ is the four--velocity of
$\Lambda_b$, and $q=p_{\Lambda_b}-p_\Lambda$ is the momentum transfer.
Comparing the general form of the form factors with (\ref{e5}), one can
easily obtain the following relations among them (see also \cite{R4814})
\bea
\label{e6}
g_1 \es f_1 = f_2^T= g_2^T = F_1 + \sqrt{r} F_2~, \nnb \\
g_2 \es f_2 = g_3 = f_3 = g_T^V = f_T^V = \frac{F_2}{m_{\Lambda_b}}~,\nnb \\
g_T^S \es f_T^S = 0 ~,\nnb \\
g_1^T \es f_1^T = \frac{F_2}{m_{\Lambda_b}} q^2~,\nnb \\
g_3^T \es \frac{F_2}{m_{\Lambda_b}} \ga m_{\Lambda_b} + m_\Lambda \dr~,\nnb \\
f_3^T \es - \frac{F_2}{m_{\Lambda_b}} \ga m_{\Lambda_b} - m_\Lambda \dr~,
\eea
where $r=m_\Lambda^2/m_{\Lambda_b}^2$. 
These relations will be used in further numerical calculations.

As has already been stated, in order to study the T--violating effects, we
need spin polarization of $\Lambda$ baryon. Spin vector $s^\mu$ of $\Lambda$
baryon can be expressed in terms of the unit vector $\vec{\xi}$ along the
$\Lambda$ spin in its rest frame as,
\bea
\label{e7}    
s_\mu = \ga \frac{\vec{p}_\Lambda \cdot \vec{\xi}}{m_\Lambda},
\vec{\xi} + \frac{\vec{p}_\Lambda (\vec{p}_\Lambda\cdot\vec{\xi})}
{m_\Lambda(E_\Lambda + m_\Lambda)}\dr~,
\eea
and choose the unit vectors along the longitudinal, transversal and normal
components of the $\Lambda$ polarization to be
\bea
\label{e8}
\vec{e}_L \es \frac{\vec{p}_\Lambda}{\vel \vec{p}_\Lambda \ver}~,\nnb \\
\vec{e}_T \es \frac{\vec{p}_{\ell^-} \times \vec{p}_\Lambda}
{\vel \vec{p}_{\ell^-} \times \vec{p}_\Lambda\ver}~,\nnb \\
\vec{e}_N \es \vec{e}_L \times \vec{e}_T~,
\eea
respectively, where $\vec{p}_{\ell^-}$ and $\vec{p}_\Lambda$ are the three
momenta of $\ell^-$ and $\Lambda$ in the center of mass frame of the final
leptons. The differential decay rate of the $\Lambda_b \rar\Lambda
\ell^+\ell^-$ decay for any spin direction $\vec{\xi}$ along $\Lambda$
baryon can be written in the following form
\bea
\label{e9}
\frac{d\Gamma}{ds} = \frac{1}{2}
\ga \frac{d\Gamma}{ds}\dr_0
\Bigg[ 1 + \Bigg( P_L \vec{e}_L + P_N
\vec{e}_N + P_T \vec{e}_T \Bigg) \cdot
\vec{\xi} \Bigg]~,
\eea
where $\ga d\Gamma/ds \dr_0$ corresponds to the unpolarized differential
decay rate, $s=q^2/m_{\Lambda_b}^2$ and
$P_L$, $P_N$ and $P_T$ represent the longitudinal, normal and
transversal polarizations of $\Lambda$, respectively.
The unpolarized decay width  in Eq. (\ref{e9}) can be written as
\bea
\label{e10}                             
\ga \frac{d \Gamma}{ds}\dr_0 = \frac{G^2 \alpha^2}{8192 \pi^5}           
\vel V_{tb} V_{ts}^\ast \ver^2 \lambda^{1/2}(1,r,s) v 
\Big[{\cal T}_0(s) +\frac{1}{3} {\cal T}_2(s) \Big]~,           
\eea                       
where 
$\lambda(1,r,s) = 1 + r^2 + s^2 - 2 r - 2 s - 2 rs$
is the triangle function and $v=\sqrt{1-4m_\ell^2/q^2}$ is the lepton
velocity. The explicit expressions for ${\cal T}_0$ and ${\cal T}_2$ can be
found in \cite{R4812}. 

The T--odd transverse polarization of $\Lambda_b$ baryon is defined as
\bea
\label{e11}
P_T(s) = \frac{\ds{\frac{d \Gamma}{ds}
                   (\vec{\xi}=\vec{e}) -
                   \frac{d \Gamma}{ds}
                   (\vec{\xi}=-\vec{e})}}
              {\ds{\frac{d \Gamma}{ds}
                   (\vec{\xi}=\vec{e}) +
                  \frac{d \Gamma}{ds}
                  (\vec{\xi}=-\vec{e})}}~.
\eea
After lengthy and straightforward calculations, we get the following
expression for $P_T(s)$

\bea
\label{e12}
\lefteqn{
P_T = - \frac{8 \pi m_{\Lambda_b}^3 v \sqrt{s\lambda}}
{\left[ {\cal T}_0 (s) +\frac{1}{3} {\cal T}_2 (s) \right]}
\Bigg\{
m_\ell \Big(
\mbox{\rm Im}[(A_1+B_1)^\ast F_1] + \mbox{\rm Im}[(A_1-B_1)^\ast H_1]     
\Big)} \nnb \\
\ar 8 m_{\Lambda_b}\Big[
 2 m_\ell \, \mbox{\rm Im}[E_1^\ast C_T f_T^V] -
(1-\sqrt{r}) \Big( \mbox{\rm Im}[H_1^\ast C_{TE} f_T] + 
2 m_\ell \, \mbox{\rm Im}[E_1^\ast C_{TE}f_T^V] \Big)\Big] \nnb \\
\ek m_\ell m_{\Lambda_b} \Big[
 - (1-\sqrt{r}) \, \mbox{\rm Im}[(A_2-B_2)^\ast H_1]+
(1+\sqrt{r}) \, \mbox{\rm Im}[(A_2+B_2)^\ast F_1] \Big] \nnb \\
\ar 8 m_\ell m_{\Lambda_b} \Big[
(1-\sqrt{r}) \, \mbox{\rm Im}[(D_3-E_3)^\ast C_T f_T] +
2 (1+\sqrt{r}) \, \mbox{\rm Im}[(D_3+E_3)^\ast C_{TE} f_T]
\Big] \nnb \\
\ek 16 m_\ell m_{\Lambda_b} \Big(
\mbox{\rm Im}[D_1^\ast (C_T+C_{TE}) f_T^V] -
\sqrt{r} \, \mbox{\rm Im}[D_1^\ast C_{TE} f_T^V] \Big) \nnb \\
\ar 8 m_\ell \Big[
\mbox{\rm Im}[(D_1-E_1)^\ast C_T f_T] -
2 \, \mbox{\rm Im}[(D_1+E_1)^\ast C_{TE} f_T] \nnb \\
\ek m_{\Lambda_b}^3 (1-\sqrt{r}) (1+ 2 \sqrt{r} +r - s)
\, \mbox{\rm Im}[(D_2-E_2)^\ast C_T f_T^S] \Big] \\
\ar m_{\Lambda_b}^2 (1-r+s) \Big(
\mbox{\rm Im}[A_2^\ast D_1 - A_1^\ast D_2] -
\mbox{\rm Im}[B_2^\ast E_1 - B_1^\ast E_2] \Big)\nnb \\
\ar 16 m_\ell m_{\Lambda_b} \Big(
\mbox{\rm Im}[(D_2-E_2)^\ast C_T f_T] +
2 \, \mbox{\rm Im}[(D_2+E_2)^\ast C_{TE} f_T] \nnb \\
\ek m_{\Lambda_b} s \, \mbox{\rm Im}[(D_3+E_3)^\ast C_{TE} f_T^V] \Big) \nnb \\
\ar 4 m_{\Lambda_b} \Big[
(1-\sqrt{r}) \, \mbox{\rm Im}[H_2^\ast C_T f_T] -
(1+\sqrt{r}) \Big( \mbox{\rm Im}[F_1^\ast C_T f_T] -
2 \, \mbox{\rm Im}[F_2^\ast C_{TE} f_T] \Big) \nnb \\
\ar m_{\Lambda_b} s  \Big( \mbox{\rm Im}[F_1 C_T f_T^V] -
2 \, \mbox{\rm Im}[F_2 C_{TE} f_T^V] \Big) \Big] \nnb \\
\ar 2 m_{\Lambda_b} \Big(
\mbox{\rm Im}[A_1^\ast E_1-B_1^\ast D_1] -
m_{\Lambda_b}^2 s \, \mbox{\rm Im}[A_2^\ast E_2 - B_2^\ast D_2] \Big) \nnb \\
\ar 8 m_\ell m_{\Lambda_b}^2 \Big(
[s-(1+\sqrt{r})^2] \, \mbox{\rm Im}[(D_1-E_1)^\ast C_T f_T^S] -
(1-r+s) \, \mbox{\rm Im}[(D_2-E_2)^\ast C_T f_T^V] \Big) \Bigg\}~.\nnb 
\eea
It follows from Eq. (\ref{e12}) that, for $P_T$ to have nonvanishing value,
\begin{itemize}
\item interactions of new type must exist, and,
\item combinations of different Wilson coefficients must have a weak phase. 
\end{itemize}
It should be noted that in the SM $P_T \sim m_s \mbox{Im} \ga C_7
C_{10}^\ast\dr$ (see \cite{R4807}) and hence it is strongly suppressed.
For this reason, if a large value for $P_T$ is measured in the experiments,
it could be considered as an unambiguous indication of the existence of new
physics beyond the SM.

It should be remembered that the electromagnetic interaction of final
particles can induce $P_T$, but this contribution is negligibly small, which
is of the order of ${\cal O}(10^{-3})$.

\section{Numerical analysis}

In this section we will investigate the sensitivity of the transversal
lepton polarization to the new Wilson coefficients. The main input
parameters in Eq. (\ref{e12}) are the transition form factors. For transition
form factors we will use the results from QCD sum rules
approach in combination with HQET \cite{R4815}, which reduces the
number of form factors into two independent ones. 
The dependence of these form factors on $s$
can be expressed as
\bea
\label{e13}
F(q^2) = \frac{F(0)}{\ds 1-a_F s + b_F s^2}~,
\eea
where parameters $F_i(0),~a$ and $b$ are listed in table 1.
\begin{table}[h]    
\renewcommand{\arraystretch}{1.5} 
\addtolength{\arraycolsep}{3pt}  
$$
\begin{array}{|l|ccc|}
\hline
& F(0) & a_F & b_F \\ \hline
F_1 &
\phantom{-}0.462 & -0.0182 & -0.000176 \\
F_2 &
-0.077 & -0.0685 &\phantom{-}0.00146 \\ \hline
\end{array}
$$
\caption{Parameters for the form factors given in Eq. (\ref{e13}) 
in the QCD sum rules method.}
\renewcommand{\arraystretch}{1}
\addtolength{\arraycolsep}{-3pt}
\end{table}

In further numerical analysis, for the values of the Wilson coefficients 
$C_7,~C_9^{eff}$, and $C_{10}$ we will use next--to leading order 
logarithmic approximation results at renormalization point $\mu=m_b$
\cite{R4816}. It should be noted that, in addition to the short distance
contribution, $C_9^{eff}$ also receives long distance contributions from the
real $\bar c c$ resonant states of the $J/\psi$ family. In the present work
we do not take into account the long distance effects.
Furthermore in carrying out numerical calculations we vary all
new Wilson coefficients in the range $0 \le C_X \le \vel C_{10} \ver$ . The
experimental bounds on the branching ratio of the $B \rar K^\ast \mu^+
\mu^-$ and $B \rar \mu^+ \mu^-$ \cite{R4817} suggest that this is the right
order of magnitude range for the vector and scalar interaction coefficients.
Moreover we assume that tensor interactions also vary in this region.
As we have already noted, in order to obtain considerably large value for
$P_T$, the new Wilson coefficients must have weak phase. For simplicity we assume
that all new Wilson coefficients have a common weak phase $\phi$. 

It follows from the explicit expression of $P_T$ that it depends on $s$,
magnitude and phase of the new Wilson coefficients. The dependence of $P_T$
is eliminated by performing integration over $s$, i.e., we analyze the
averaged $P_T$, which is defined as
\bea
\label{e14} 
\lla P_T \rra = \ds\frac{\int_{4
m_{\ell}^2/m_{\Lambda_b}^2}^{(1-m_\Lambda/m_{\Lambda_b})^2} 
P_T \ds\frac{d{\cal B}}{ds} ds}{
\int_{4 m_{\ell}^2/m_{\Lambda_b}^2}^{(1-m_\Lambda/m_{\Lambda_b})^2} 
\ds\frac{d{\cal B}}{ds} ds}~.
\eea 

The dependence of the averaged $\lla P_T \rra$ on the Wilson coefficients 
$C_T$, $C_{TE}$, $C_{RL}$, $C_{RR}$, $C_{RLRL}$ and $C_{RLLR}$ and on the 
phase $\phi$ for the $\Lambda_b \rar \Lambda \tau^+ \tau^-$ decay is
presented in Figs. (1)--(6).  
The dependence of $\lla P_T \rra$ on $C_{RL}$
and on the weak phase for the $\Lambda_b \rar \Lambda \mu^+ \mu^-$ decay is
depicted in Fig. (7). On the other hand, the dependence of 
$\lla P_T \rra$ on the remaining Wilson coefficients 
$C_{LL},~C_{LR},~C_{LRRL}$ and $C_{LRLR}$ for the $\Lambda_b \rar \Lambda
\mu^+ \mu^-$ decay is not presented since for
all these choices $\lla P_T \rra$ is negligibly small ($<0.2\%$). 

It can easily be seen from these figures that $\lla P_T \rra$ attains at its
largest value for $C_{RL}$ about $\sim 25\%$, for
$C_{RR}\sim 4\%$, for $C_{TE}\sim 8\%$, for $C_{RLRL}$, $C_{RLLR}$ about
$\sim 5\%$ and about $2\%$ for $C_T$ for the $\Lambda_b \rar \Lambda \tau^+
\tau^-$ decay.

The situation for the $\Lambda_b \rar
\Lambda \mu^+ \mu^-$ decay is similar to the previous case. It is
observed that $\lla P_T\rra$ displays even stronger dependence only on $C_{RL}$   
and reaches its maximum value of $\sim 40\%$, and for $C_T$ about $1\%$,
while for all remaining Wilson coefficients other than these two $\lla
P_T\rra$ is negligibly small ($\simlt 0.2\%$).
    
These behaviors can be explained as follows. In the massless lepton limit 
(neglecting scalar type interactions for simplicity)
the expression for $P_T$ reduces to
\bea
\label{e15}
P_T &\sim& \frac{1}{{\cal T}_0 + \frac{1}{3} {\cal T}_2}
\Big\{ A \, \mbox{\rm Im}[C_{BR}^\ast \ga C_{RR}-C_{RL}\dr - C_{SL}^\ast \ga
2 C_{10}+C_{LR}-C_{LL}\dr \Big] \nnb \\
\ar B \, \mbox{\rm Im}[\ga C_{RL}^\ast+C_{RR}^\ast\dr \ga 
2 C_{10}+C_{LR}-C_{LL}\dr + \ga 2 C_9 + C_{LR}+C_{LL}\dr
\ga C_{RR}^\ast -C_{RL}^\ast\dr ]\Big\}~, 
\eea
where $A$ and $B$ are functions of $s$ and ratio of the form factors.
In this work we will assume that $C_{BR}$ and $C_{SL}$ are defined in the
same way as in the SM, since the measured branching ratio of $B \rar X_s \gamma$ decay
is in a very good agreement with the SM prediction \cite{R4818}--\cite{R4820}.     
The first term in Eq. (\ref{e15}) can safely be neglected since $\vel
C_7^{SM} \ver \ll \vel C_9^{eff} \ver$ and $\vel C_7^{SM} \ver \ll \vel
C_{10} \ver$. We also observe that the
expression in Eq. (\ref{e15}) involves $C_{RR}$ and $C_{RL}$ in the
combinations of the form
\bea
&&\sim \mbox{\rm Im}[C_{RR}^\ast \ga C_{10} + C_9^{eff} \dr]~,\nnb \\
&&\sim \mbox{\rm Im}[C_{RL}^\ast \ga C_{10} - C_9^{eff} \dr]~.\nnb
\eea
In the SM $C_{10}$ and $C_9^{eff}$ have opposite signs and have almost
equal magnitudes (see for example \cite{R4821}),  
which explains the reason why $P_T$ gets larger values for $C_{RL}$
compared to the other new Wilson coefficients. For all other choices $P_T$
is proportional to the lepton mass and for this reason $P_T$ is quite small
the coefficients $C_{LL},~C_{LR},C_{LRLR}$ and $C_{LRRL}$. 

Finally we would like to discuss briefly the detectibility of $\lla P_T
\rra$ in the experiments. Experimentally, to measure $\lla P_T \rra$ of a 
particular decay with the branching ratio ${\cal B}$ at the $n
\sigma$ level, the required number of events are $N=n^2/({\cal B} \lla P_T
\rra )^2$. For example, if $B(\Lambda_b \rar \Lambda \mu^+ \mu^-) \sim
10^{-6}$ then, to measure $\lla P_T\rra\simeq 0.2$ at $3 \sigma$ level,
$N\simeq 4.5\times 10^7$ $\Lambda_b$ decays are required. Since at LHC and
BTeV machines $10^{12}~\bar b b$ pairs are expected to be produced per year
\cite{r4822},
the detectibility of $\lla P_T\rra$ is quite high in these colliders.

\newpage

\section*{Figure captions}
{\bf Fig. (1)} The dependence of the averaged transversal lepton
polarization $\lla P_T^- \rra$ on the phase $\phi$ and on the new Wilson 
coefficient $C_T$ for the $\Lambda_b \rar \Lambda \tau^+ \tau^-$ decay.\\\\
{\bf Fig. (2)} The same as in Fig. (1), but for the Wilson coefficient
$C_{TE}$.\\\\
{\bf Fig. (3)} The same as in Fig. (1), but for the Wilson coefficient
$C_{RL}$.\\\\
{\bf Fig. (4)} The same as in Fig. (1), but for the Wilson coefficient
$C_{RR}$.\\\\
{\bf Fig. (5)} The same as in Fig. (1), but for the Wilson coefficient
$C_{RLRL}$.\\\\
{\bf Fig. (6)} The same as in Fig. (1), but for the Wilson coefficient
$C_{RLLR}$.\\\\
{\bf Fig. (7)}  The dependence of the averaged transversal lepton
polarization $\lla P_T^- \rra$ on the phase $\phi$ and on the new Wilson 
coefficient $C_{RL}$ for the $\Lambda_b \rar \Lambda \mu^+ \mu^-$ decay.

\newpage

\begin{figure}  
\vskip 1.5 cm   
    \includegraphics{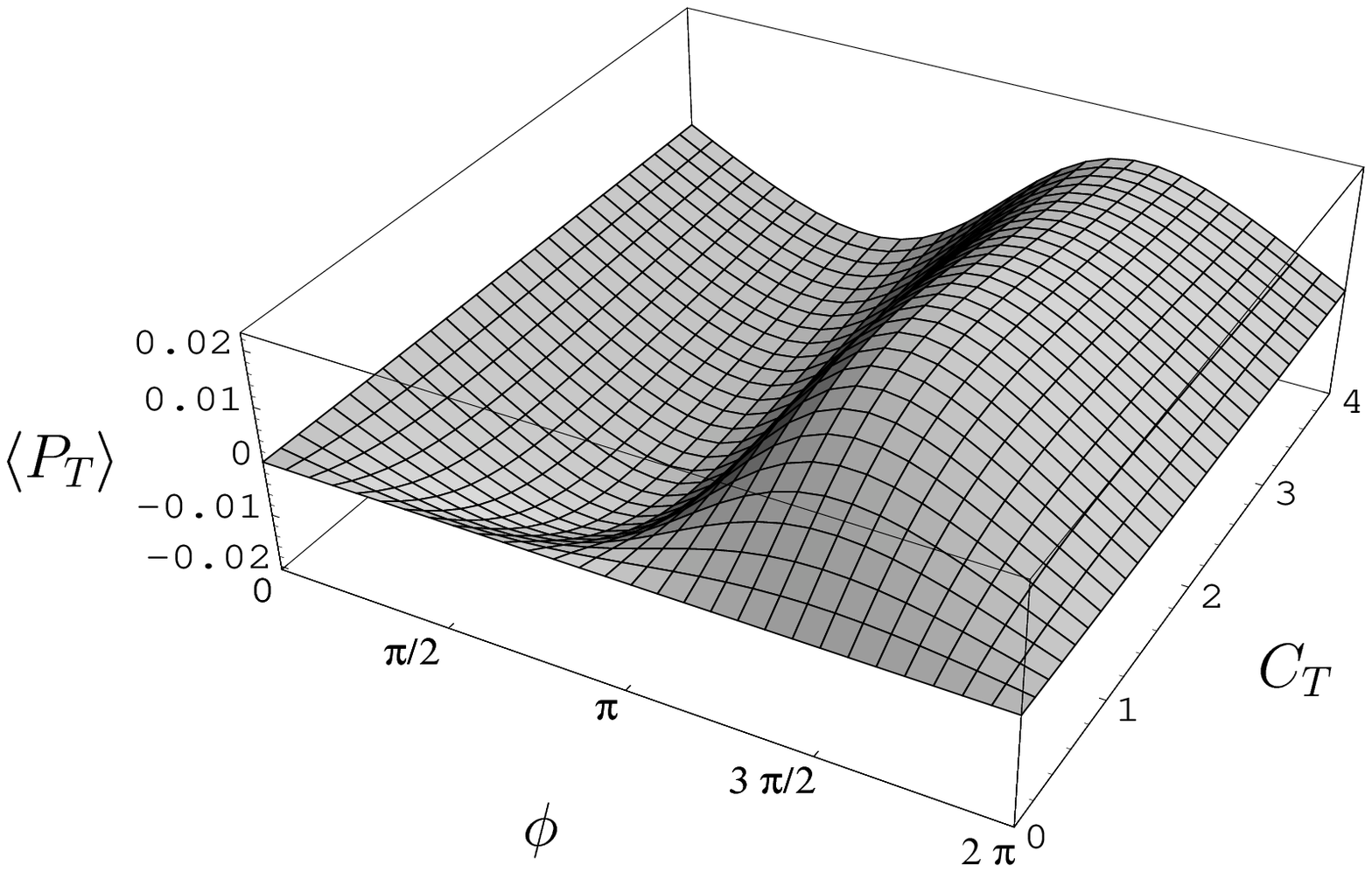}
\vskip 7.0cm     
\caption{}
\end{figure}

\begin{figure}
\vskip 1. cm
    \includegraphics{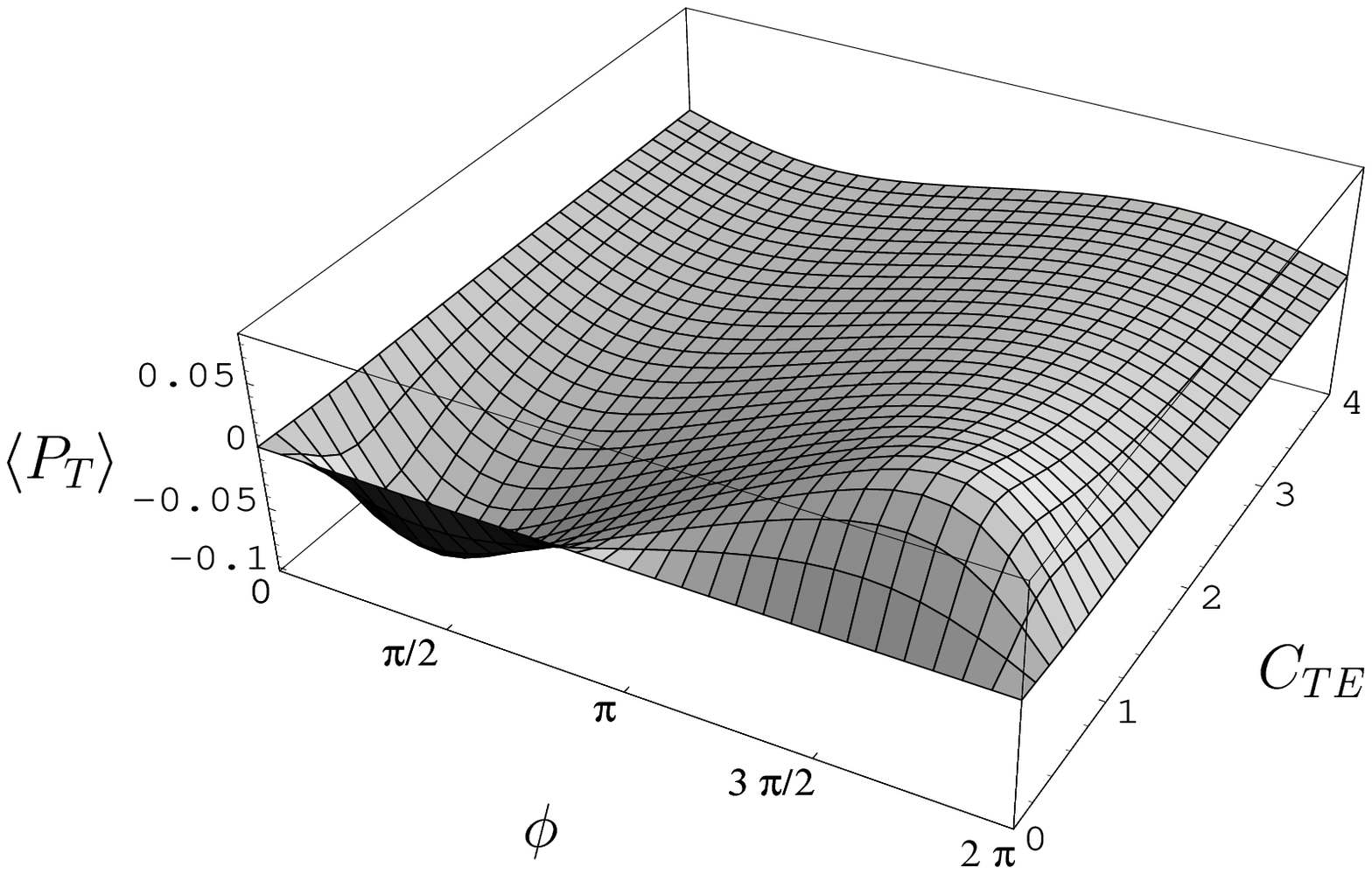}
\vskip 8.0 cm
\caption{}
\end{figure}

\begin{figure}  
\vskip 1.5 cm   
    \includegraphics{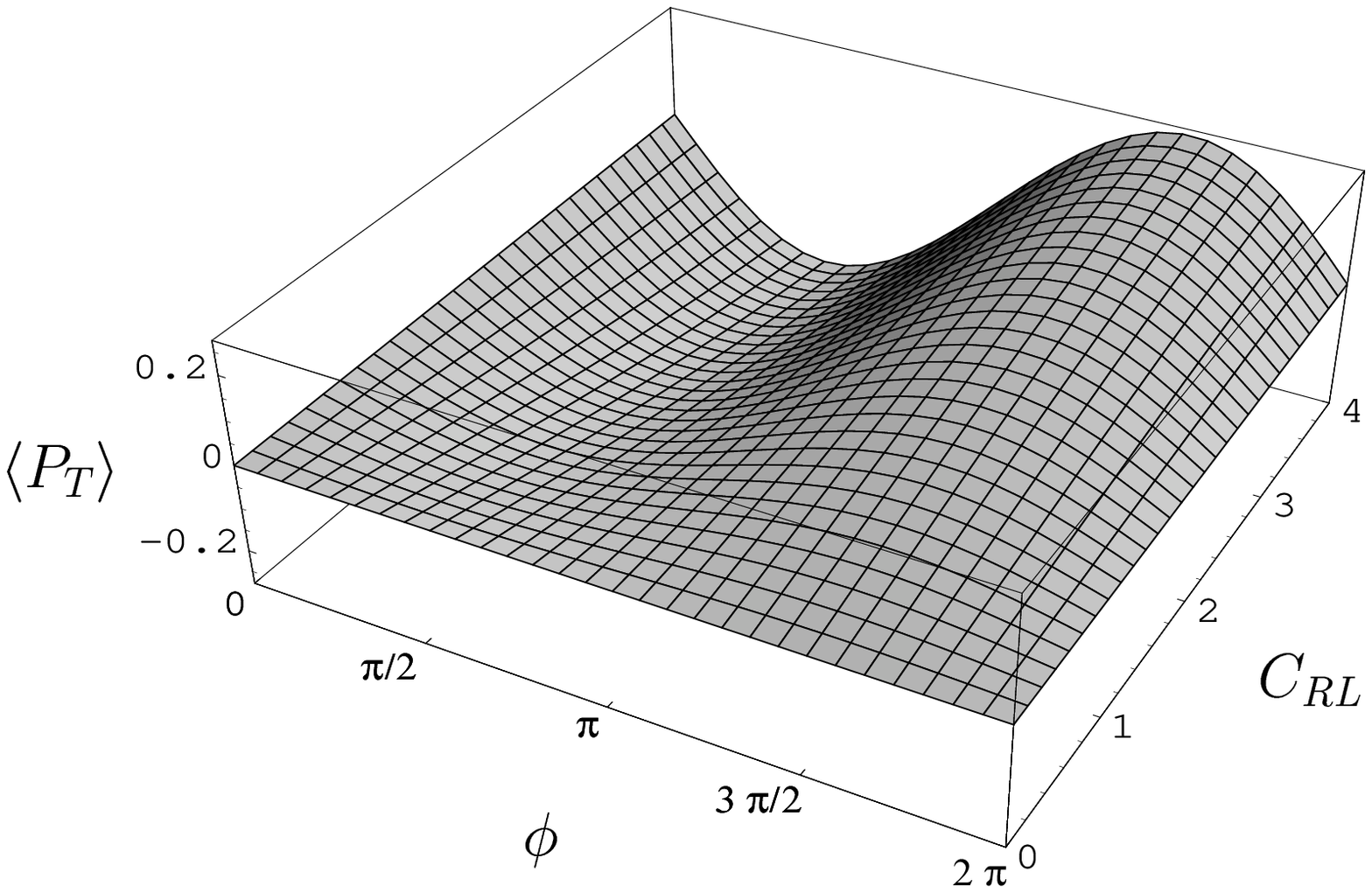}
\vskip 7.0cm     
\caption{}
\end{figure}

\begin{figure}
\vskip 1. cm
    \includegraphics{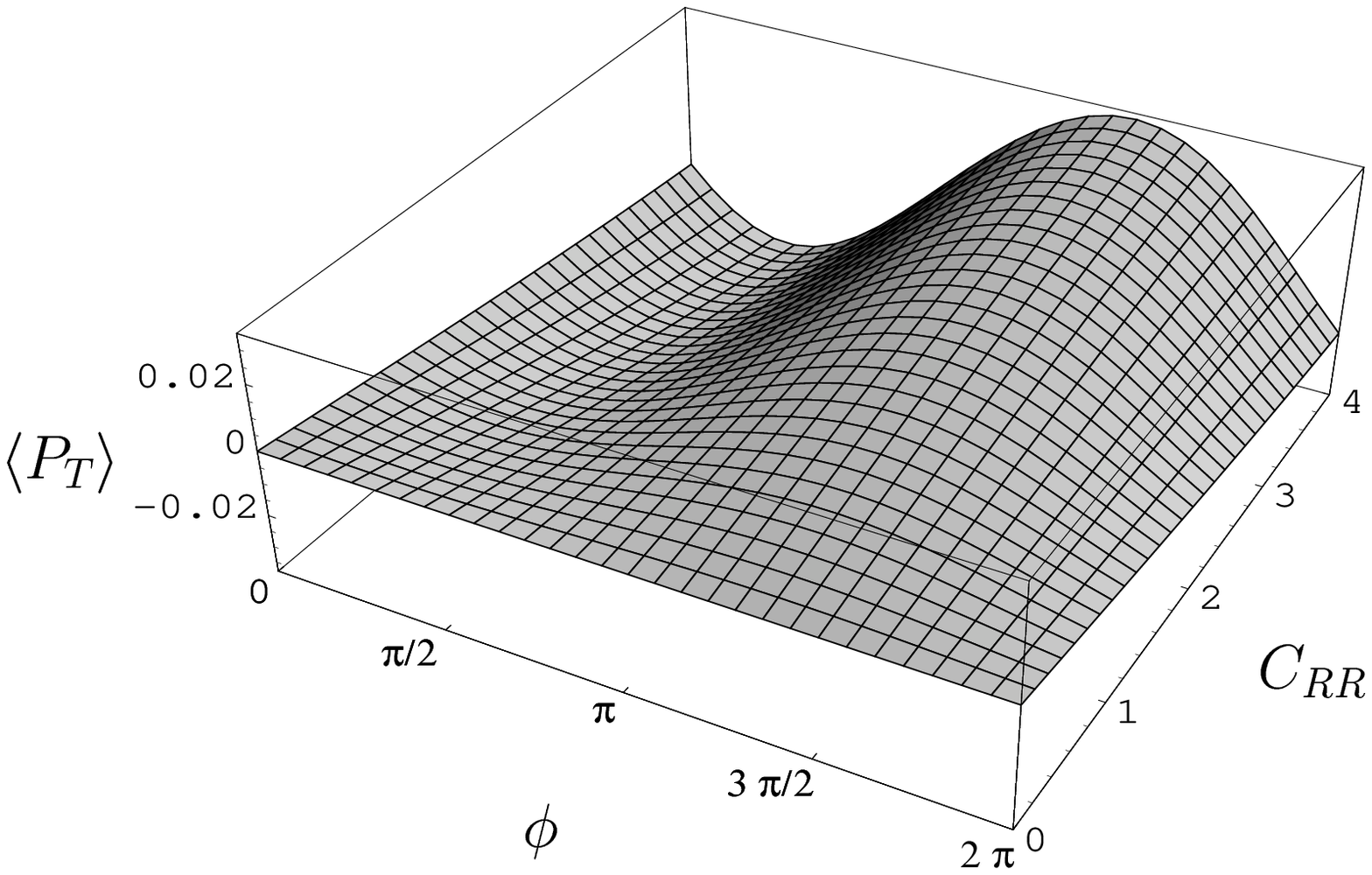}
\vskip 8.0 cm
\caption{}
\end{figure}

\begin{figure}  
\vskip 1.5 cm   
    \includegraphics{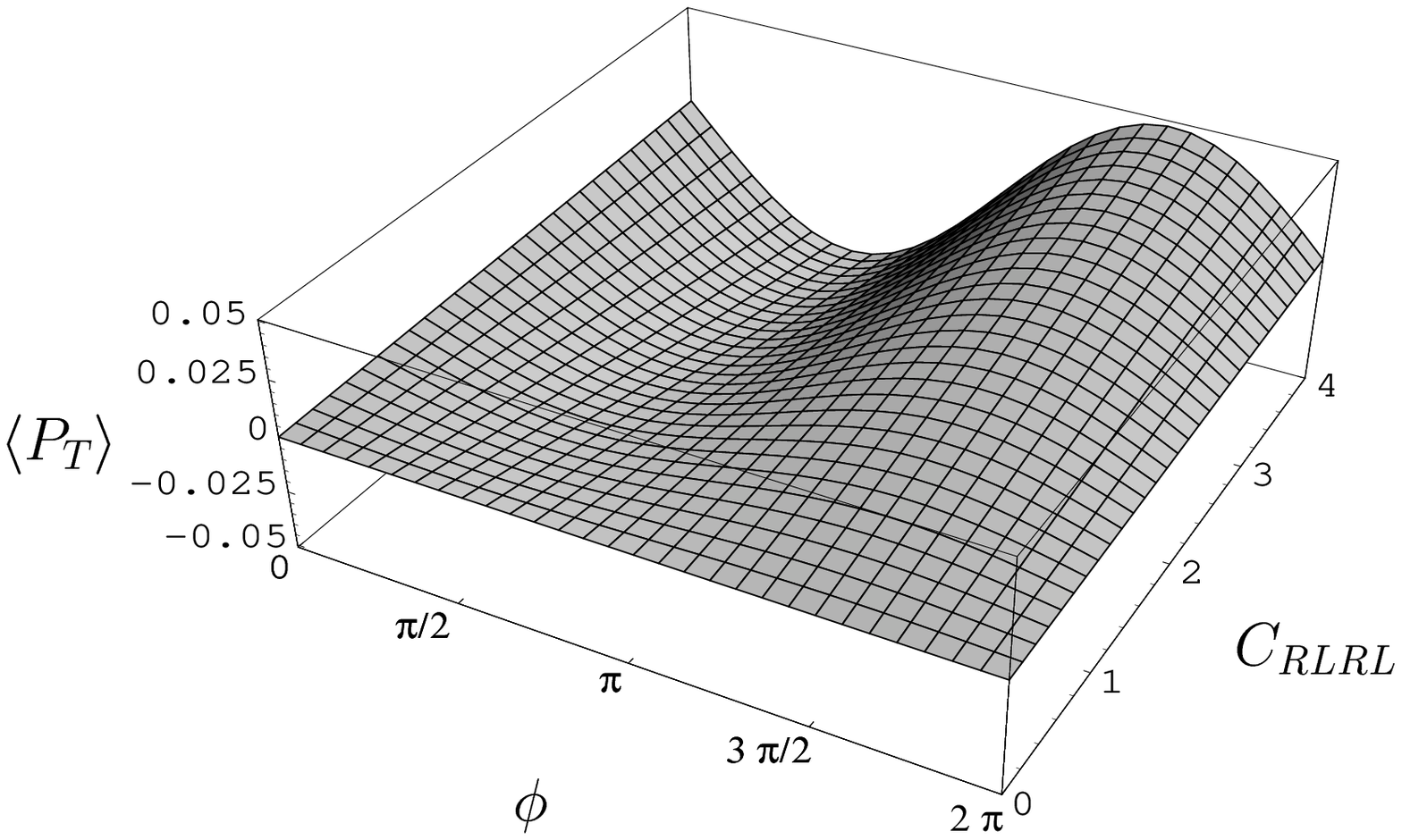}
\vskip 7.0cm     
\caption{}
\end{figure}

\begin{figure}
\vskip 1. cm
    \includegraphics{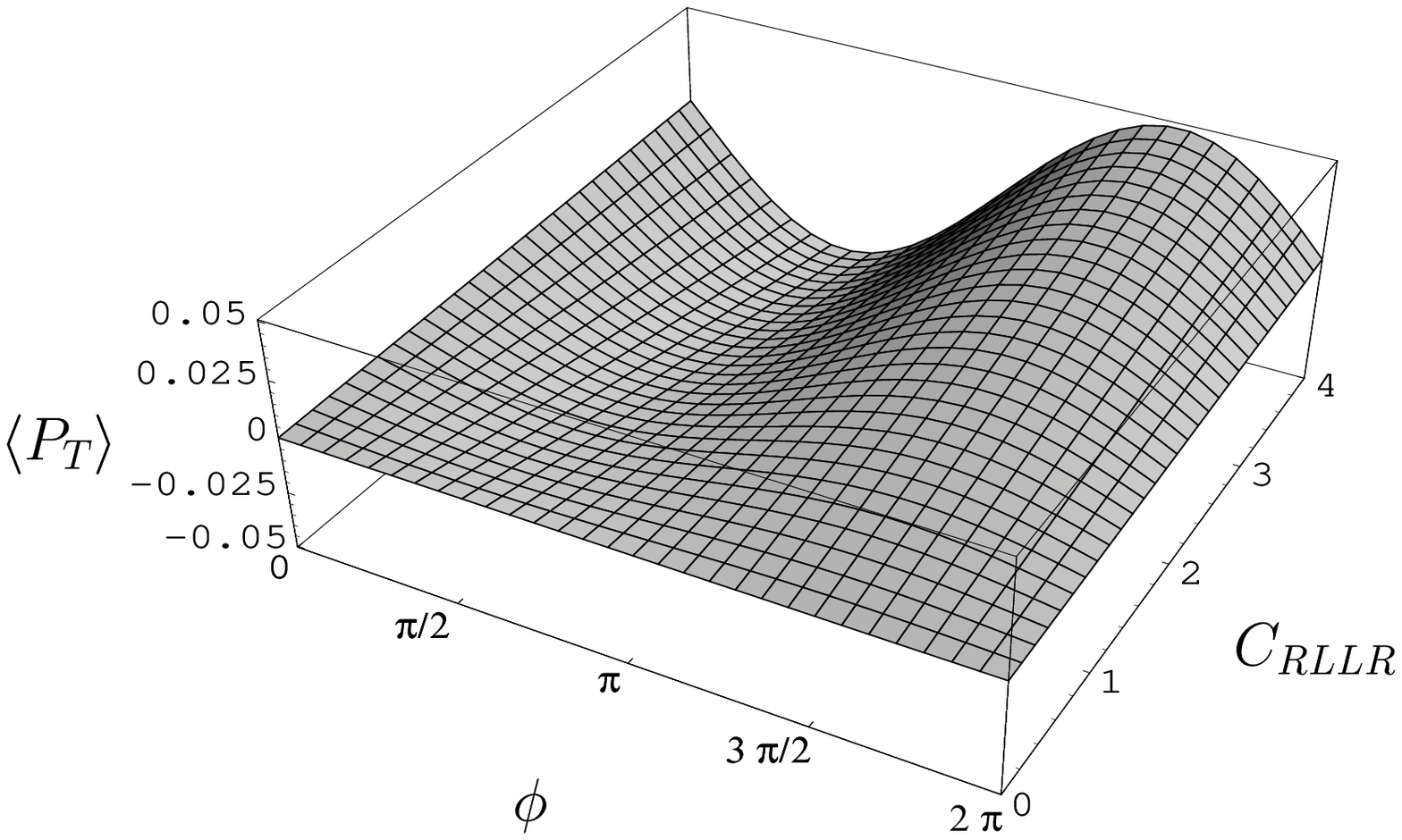}
\vskip 8.0 cm
\caption{}
\end{figure}

\begin{figure}  
\vskip 1.5 cm   
    \includegraphics{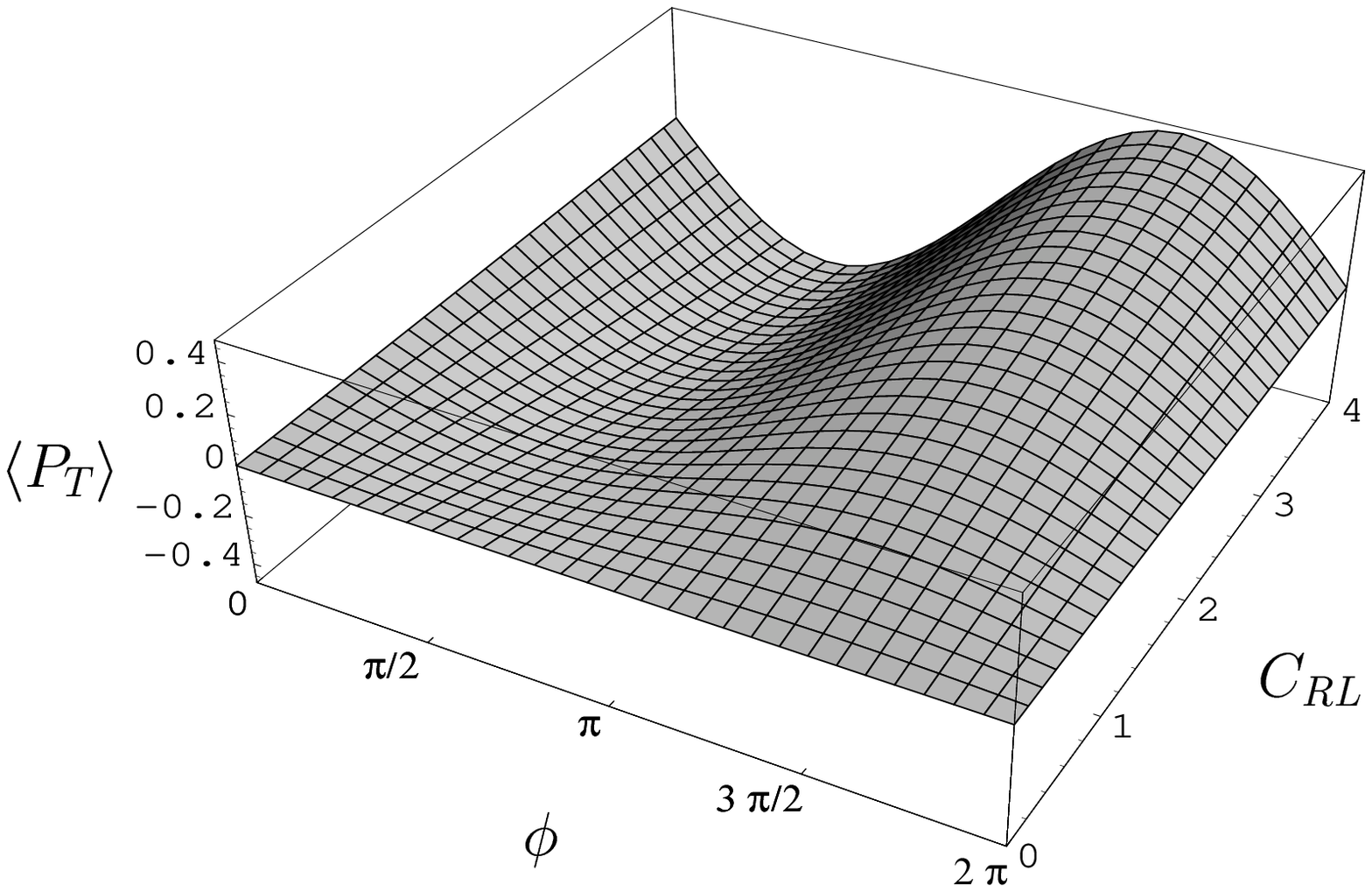}
\vskip 7.0cm     
\caption{}
\end{figure}

\end{document}